\documentstyle{article}
\newcommand{\be}{\begin{equation}}
\newcommand{\ee}{\end{equation}}
\newcommand{\mb}{\mbox}
\newcommand{\pr}{\prime}
\newcommand{\p}{\partial}
\newcommand{\fr}{\frac}
\newcommand{\rar}{\rightarrow}

\begin{document}
 
\title{Averaging in cosmology}
\author{Jelle P. Boersma \\ ~ \\
Department of Applied Mathematics, University of Cape Town \\
Rondebosch 7700, Cape Town, South Africa}
\maketitle
\begin{abstract}
In this paper we discuss the effect of local inhomogeneities
on the global expansion of nearly FLRW universes,
in a perturbative setting.
We derive a generic linearized averaging operation for metric
perturbations from basic assumptions, and we explicify the
issue of gauge invariance.
We derive a gauge invariant expression for the back-reaction of
density inhomogeneities on the global expansion of perturbed
FLRW spacetimes, in terms of observable quantities,
and we calculate the effect quantitatively.
Since we do not adopt a comoving gauge, our result incorporates
the back-reaction on the metric due to scalar velocity and
vorticity perturbations.
The results are compared with the results by other authors
in this field.
 
\end{abstract}

\section{Introduction}
 
An essential difficulty which occurs when dealing with realistic cosmological
models, is related to the fact that although the universe seems to be
very close to FLRW at length scales of the order of the Hubble radius,
the metric and matter content of the universe appears to be highly
inhomogeneous at smaller scales.
Since the realistic universe, with all its details at length scales
small compared to the Hubble radius, is too complicated to
handle in most calculations, it seems desirable to extract
those physical quantities which describe the large scale
structure of the universe.
However, when one tries to define an averaging operation for
metrics, a number of difficulties occur.
One of these difficulties is related to the fact that the Einstein
equations are inherently nonlinear, which makes it a nontrivial
question to see how the Einstein equations constrain the
dynamics of an averaged metric.
Another fundamental problem which occurs when one tries to average
metrics, is related to the fact that there is generally no
direct physical significance in an averaged metric.
Although this problem is usually ignored in the literature
on averaging, it needs to be addressed before one can extend
the discussion on averaging beyond an intuitive level
of understanding.
The usual approach to averaging (see e.g., \cite{futamase} - \cite{mauro2})
seems to be that one defines an averaging method, which is
chosen on the basis of mathematical elegance or an intuitive
notion of smoothness, and then one defines averaged
physical quantities by means of the averaging
operation which one has chosen.
The objection against this approach is that if one calculates e.g. the averaged
expansion of a perturbed FLRW universe, one can obtain virtually any
result, by choosing an averaging operation which
yields this specific result.
In section \ref{fff} we explicify this problem, and we derive
a generic linearized averaging operation for {\em metrics}
which satisfies the condition that unperturbed FLRW is a {\em stable fixed point}
of the averaging operation.
It is shown that this generic linearized averaging operation for metrics can
be expressed in terms of the spatial average of the perturbation of the
spatial volume and $g_{00}$ in coordinates which are synchronous
in the background.
In section \ref{gp} we discuss the
gauge problem and the choice of the
background spacetime.
The averaged constraint equations are explicitly evaluated
in section \ref{ace}, and in subsection \ref{asc} we derive
an expression for the correction to averaged expansion due
to density perturbations,
in terms of the power spectrum of the matter.
In subsections \ref{aed} and \ref{sqs} we discuss the back-reaction
on the metric due to matter velocity perturbations, and we show
that vorticity perturbations may be important
in the large wavelength limit.
In section \ref{rdn} we calculate the different corrections
to the averaged expansion quantitatively by means of the
observational data, and we compare
our results with the results derived in previous works.
In this paper, we adopt the convention that greek indices
run from 0 to 3, while latin indices run from 1 to 3.
The metric signature is $(- + + +)$, and the velocity
of light, $c$, is set equal to one.

\section{The spatial average}\label{fff}
 
In this section we will consider the generic linearized averaging operation
for metrics, for which unperturbed FLRW is a stable fixed point,
and we show that this averaging operation has a universal
limit when applied iteratively to perturbed
FLRW spacetimes.
We determine this limit explicitly, and by using the symmetry of the
background, we show that the averaging of the 10 components of
the metric perturbation $\delta g_{\rho\sigma}$, can be expressed
in terms of the spatial average of $g_{00}$ and $\sqrt{g^{(3)}}$
in coordinates which are synchronous in the background.
>From now on the background FLRW spacetime will be called $\bar{\bf S}$,
while the inhomogeneous spacetime is called ${\bf S}$.
Furthermore, we assume that $\bar{\bf S}$ is coordinatized such that
$t$ represents the time coordinate which labels the hypersurfaces
of homogeneity $\bar{\Sigma}$ in $\bar{\bf S}$, and $\bar{\Sigma}$
is coordinatized by $x^i$ where $i \in \{1,2,3 \}$.
We call a metric $g_{\mu\nu}$ or a metric perturbation $\delta g_{\mu\nu}$
spatially homogeneous and isotropic when there exists
at least one coordinate system in which the components of
$g_{\mu\nu}$ or $\delta g_{\mu\nu}$ are spatially homogeneous
and invariant under spatial rotations.
 
Let us consider the most general averaging operation $\hat{A}$,
which is a functional ${\cal F}$ of metric perturbations
$\delta g_{\mu \nu}$
about some background solution $\bar{\bf S}$,
\be\label{genav}
\  \hat{A} \delta g_{\mu \nu} (x)
\ = {\cal F}_{\mu\nu} (\delta g_{\rho\sigma} (x) ).
\ee
We will {\em require} the condition that unperturbed FLRW, in
a gauge where the metric perturbation $\delta g_{\mu \nu}$ is
spatially homogeneous and isotropic, is a stable fixed point
of the averaging operation $\hat{A}$.
This condition states that the averaging operation
increases the spatial symmetry of the spacetime on which it works,
assuming that this spacetime is sufficiently `close'
to FLRW, and it defines what we mean by averaging
in this paper.
 
It follows directly from this assumption that
\be\label{c1}
\ {\cal F}_{\mu\nu} (0) = 0,
\ee
for all $x$, since a nonzero value at the right-hand side
of equation (\ref{c1}) implies that unperturbed FLRW with the
the same geometry as $\bar{\bf S}$, in a gauge where
$\delta g_{\mu \nu} =0$ for all $x$, is not a
fixed point of $\hat{A}$, which contradicts
our assumption.
 
The {\em linear approximation} to the averaging operation (\ref{genav}),
is given by
\be\label{spav}
\ \hat{A}^{(1)} \delta{g}_{\mu\nu} (x) =
\ \int_{\bar{S}} {\it d}^4 x^{\pr}
\   f^{\rho\sigma}_{\mu\nu}
\ (x,x^{\pr}) \delta{g}_{\rho\sigma} (x^{\pr}) ,
\ee
where the bi-tensor density
$f^{\rho\sigma}_{\mu\nu} (x,x^{\pr})$ is defined as the functional
derivative of ${\cal F}_{\mu\nu}$ with respect to $\delta g_{\rho\sigma}$,
evaluated at the point with coordinates $x^{\pr}$ in the
background, i.e.,
\be\label{f}
\ f^{\rho\sigma}_{\mu\nu} (x,x^{\pr}) :=
\ \left. \fr{\p {\cal F}_{\mu\nu} (g,x) }{\p g_{\rho\sigma} (p)}
\ \right|_{\delta g_{\rho\sigma} =0,p=x^{\pr} },
\ee
and we used condition (\ref{c1}) which states that the zeroth
order contribution in the expansion of $\hat{A}$ vanishes.
 
The condition that unperturbed FLRW is a {\em stable} fixed
point of the averaging operation $\hat{A}$ implies that
the limit
\be\label{c2}
\ \hat{A}^{\infty} \delta g_{\mu \nu}
\ := \lim_{n \rightarrow \infty} \hat{A}^{(1) n} \delta g_{\mu \nu}
\ee
exists, and the quantity
$\hat{A}^{\infty} \delta g_{\mu \nu}$ must be spatially
homogeneous and isotropic (we used the notation
$\hat{A}^{(1) n}$ to denote the $n$-times repeated operation
of $\hat{A}^{(1)}$).
 
Note that the averaging operation $\hat{A}$ has two aspects; first it
changes the geometry of the  spacetime on which it works, and second
it specifies a correspondence between points in the spacetime ${\bf S}$,
the averaged spacetime $\hat{A}{\bf S}$, and the background
$\bar{\bf S}$.
 
When one only requires that unperturbed FLRW is a stable fixed point of
$\hat{A}$, one constrains the way in which $\hat{A}$ changes the
geometry of the spacetime on which it works, but one does not constrain
the correspondence between points in the spacetimes ${\bf S}, \hat{A}{\bf S}$
and $\bar{\bf S}$.
We constrain this freedom by imposing the stronger requirement
that unperturbed FLRW, in a gauge where the metric perturbation
$\delta g_{\mu \nu}$ is spatially homogeneous and isotropic,
is a stable fixed point of $\hat{A}$.
This condition enforces that $\hat{A}$ does not generate `pure gauge'
perturbations when operating on unperturbed FLRW.
 
Starting from equation (\ref{c2}), and using the symmetries
of the background spacetime $\bar {\bf S}$, it is shown in
appendix B that the averaging operation
$\hat{A}^{\infty}$ can be defined in terms of
a spatial averaging operation which is universal, i.e.,
\be\label{c2X}
\ \hat{A}^{\infty} \delta g_{\mu \nu} (t,x^i)
\ = \langle \delta g_{\mu \nu} \rangle (t),
\ee
where
\be\label{hddk}
\ \langle \delta g_{\mu\nu} \rangle (t)=
\ \int_{\bar{\Sigma} (t)} {\it d}^3 x^{\pr} \alpha \sqrt{{g}^{(3)}}
\ee
\[
\ \times \left[
\ \bar{n}^{\rho} (x^{\pr})  \bar{n}^{\sigma} (x^{\pr})
\ \bar{n}_{\mu} (x) \bar{n}_{\nu} (x)
\ + \fr{1}{3}  \bar{h}^{\rho\sigma} (x^{\pr}) \bar{h}_{\mu\nu} (x) \right]
\ \delta g_{\rho\sigma}  (x^{\pr}) ,
\]
where $\bar{n}^{\rho}$ denotes the future directed unit vector normal
to $\bar{\Sigma}$, and
$\bar{h}^{\rho\sigma} := g^{B \rho\sigma} + \bar{n}^{\rho} \bar{n}^{\sigma}$
is the projection operator on $\bar{\Sigma}$, and $\alpha$ denotes
the distribution which is constant
on ${\Sigma}$, and for which the integral over
${\Sigma}$ equals one.
Note that $ \bar{n}^{\rho} \bar{n}^{\sigma}  \delta g_{\rho\sigma}$
equals the perturbation of $g_{00}$ in coordinates which are synchronous
in the background (i.e., coordinates for which
$ {g}^B_{\mu 0} = - \delta_{\mu}^0$),
while $\bar{h}^{\rho\sigma}\delta g_{\rho\sigma}$ equals the perturbation
of the spatial volume element on $\Sigma$, to first
order.
It follows from this observation that the linearized averaging
operation for metrics (\ref{hddk}), is effectively a spatial
averaging operation for scalars, applied to $\delta g_{00}$
and $\delta g^i_{\;i}$ in coordinates which are synchronous in
the background.
 
An explicit realization of the spatial averaging
operation for a scalar $q (x)$, in the
case where $\bar{\Sigma}$ is open, is given by
\be\label{avinfty}
\ \langle q (x) \rangle =
\  \lim_{\ell \rar \infty} \langle q (x) \rangle (\ell)
\ee
\[
\ := \lim_{\ell \rar \infty}
\  N^{-1} (x,\ell)
\ \int_{\bar{\Sigma}} {\it d}^3 x^{\pr} \sqrt{g^{ (3)}}
\  q (x^{\pr}) \theta ( \ell - \Delta s (x,x^{\pr}) ) ,
\]
where
$
\ N (x,\ell) :=  \int_{\bar{\Sigma}} {\it d}^3 x^{\pr}   \sqrt{g^{(3)}}
\  \theta ( \ell - \Delta s (x,x^{\pr}) ),
$
and
$
\  \Delta s (x,x^{\pr})
$
is a distance measure between points $x$ and $x^{\pr}$, $\ell$ is a parameter
with the dimension of length,
and $\theta (x) = 1 (0)$ for $x \geq 0 ( x < 0)$.
In the case where ${\Sigma}$ is closed, $\langle q \rangle$
is defined analogously to expression (\ref{avinfty}),
with $N (x,\ell) =$volume $({\Sigma})$.
 
It is shown in appendix A that the spatial average of a scalar function
is invariant under {\em spatial} gauge transformations, to arbitrary order
in the expansion parameter of the gauge transformation.

Notice that the spatial average (\ref{avinfty})
is only well defined when we make the assumption
that perturbations $q (x)$ are sufficiently small, such that
the limit $\ell \rar \infty$ in equation
(\ref{avinfty}) exists.
It should be stressed that this assumption is nontrivial,
and it is not automatically satisfied in general
cosmological situations, where perturbations are not
necessarily bounded in amplitude and length
scale.
Indeed, since the observable part of our universe is restricted
to our past light cone, there is no observational basis for
the assumption that our universe is 'close' to FLRW at
arbitrary large length scales.
The usual way to deal with this situation is that one adopts {\em a priori}
philosophical assumptions, such as the Copernican principle,
to choose between different models which  satisfy the
observational data (see e.g., \cite{ellis}).
Throughout this paper, we will adopt
a version of the Copernican principle by assuming
that perturbations are small enough
such that the limit
$\ell \rar \infty$ in equation (\ref{avinfty}) exists.

\section{The gauge problem}\label{gp}

As is pointed out by Futamase in \cite{futamase}, the observed matter density
contrasts are of the order of unity at dimensionless length scales $\kappa$
of the order of $10^{-2}$, where $\kappa$ denotes the fraction of typical
size of the density fluctuation and the Hubble radius $r_H := c/ H_0$.
A rough estimation of the order of magnitude of the associated
Newtonian gravitational potential,
which we call $\epsilon$ from now on, can be obtained by using the
Poisson equation. For density contrasts of the order of unity
we find $\epsilon \sim \kappa^{2}$, which implies a Newtonian
potential $\phi$ of the order of $10^{-4}$, suggesting that a
perturbative approach might be adequate.
At length scales of the order of the Hubble radius, the observable
part of the universe appears to be highly homogeneous and isotropic,
which motivates our choice for the FLRW metric as a background
metric.
 
Let us first briefly discuss some details concerning
the spherical harmonic decomposition of perturbations
about a background FLRW spacetime.

The FLRW background metric can be written in the form,
\be\label{bg}
\ {\it d}s^2 = g^B_{\mu\nu} {\it d} x^{\mu} {\it d} x^{\nu} =
\ a^2 (\bar{t}) ( - {\it d} \bar{t}^2 +  \eta_{ij} {\it d} x^i {\it d} x^j),
 \ee
where $\eta_{ij}$ is the metric tensor for a homogeneous and isotropic
three-space with curvature $\bf{k}$, and $\bar{t}$ is a
{\em conformally scaled} time parameter.
We define the metric perturbation $h_{\mu\nu}$ by,
\be
\ g_{\mu\nu} = g^{B}_{\mu\nu} + h_{\mu\nu}\;\;\;,\;\;\;\;
\ g^{\mu\nu} = g^{B \mu\nu} - h^{\mu\nu},
\ee
and since $g^{\mu\rho} g_{\mu\nu} =\delta_{\; \nu}^{\rho}$, we have
$h^{\mu\nu} = g^{B \nu \rho} g^{B \mu \sigma}  h_{\rho\sigma}$, and
$h^{\mu}_{\;\nu} = g^{B \mu\rho} h_{\rho\nu}$,
to first order.
 
Copying Bardeen's notation in \cite{bardeen}, we define scalar, vector and
tensor spherical harmonics $Q^{(0)}_n,Q^{(1)}_{n\;i}$ and $Q^{(2)}_{n\;ij}$,
respectively, which satisfy the Helmholtz equations
$Q^{(p) |q}_{n \;\;\;\,|q} + k_n^2 Q^{(p)}_n = 0$, where $p \in \{0,1,2 \}$
and $|$ denotes the covariant derivative with respect to
$g^B_{i j}$. The vector harmonics $Q^{(1)}_i$ are divergenceless,
while the tensor harmonics $Q^{(2)}_{ij}$ are divergenceless,
symmetric, and traceless.
We define traceless symmetric scalar harmonics $Q^{(0)}_{n\, ij}$ by
\be\label{eee0}
\ Q^{(0)}_{n\, ij}:= k_n^{-2} Q^{(0)}_{n |ij}+ \fr{1}{3} g^{B}_{ij} Q^{(0)}_n
\ \;\; \mb{and} \;\;
\ Q^{(0) ij}_{n \;\;\;\,|ij} -  (k_n^2 - 3 {\bf k})  Q^{(0)}_n = 0,
\ee
and traceless symmetric vector harmonics $Q^{(1)}_{n\, ij}$ are defined by
\be\label{eee1}
\ Q^{(1)}_{n\, ij} :=  -  \fr{1}{2 k_n} ( Q^{(1)}_{n i|j} + Q^{(1)}_{n j|i} )
\ \;\;\mb{and}\;\;
\ Q^{(1) |i }_{n\, ij} - (k_n^2 - 2{\bf k}) Q^{(1)}_{n j} =0.
\ee
 
The spherical harmonics are labeled by the parameter
$n \in 0,{\bf Z}^+$ ($\vec{k} \in {\bf R}^3$) in the case where
$\bar{\Sigma}$ is closed (open).
It is useful to define the hypersurface integration operation for scalars
$q (x)$ by
\be\label{hypint}
\   \langle \langle q  \rangle \rangle :=  \langle q
\ ( {g}^{(3) B}/ g^{(3)})^{\fr{1}{2}} \rangle ,
\ee
which differs from the spatial average (\ref{avinfty})
by the volume element which is evaluated in the background.
As we show in appendix 2, the spatial average (\ref{avinfty}) of a physical
quantity is invariant under spatial gauge transformations,
while the hypersurface integral (\ref{hypint}) is generally gauge dependent
at second and higher order in the expansion parameter of the gauge
transformation.

 The spherical harmonics $Q^{(0)}_n, Q^{(1)}_{n\,i}$ and $Q^{(2)}_{n\,ij}$ are
orthogonal with respect to the hypersurface integration
operation, i.e.,
\be\label{nrm}
\  \langle \langle Q^{(0)}_n Q^{(0)}_{n^{\pr}}\rangle \rangle =
\  \langle \langle Q^{(1)}_{n \,i} Q^{(1) i}_{n^{\pr}}\rangle \rangle =
\   \langle \langle Q^{(2)}_{n \, ij} Q^{(2) ij}_{n^{\pr}} \rangle \rangle
\ = \delta_{n n^{\pr}},
\ee
and
\be\label{q3}
\ \fr{3}{2} \langle \langle Q^{(0)}_{n ij} Q^{(0) ij}_{n^{\pr}} \rangle \rangle
\ = 2  \langle \langle Q^{(1)}_{n \, ij} Q^{(1) ij}_{n^{\pr}} \rangle \rangle
\  =  \delta_{n n^{\pr}}.
\ee
Notice that the spherical harmonics are only to zeroth order
orthogonal with respect to the spatial averaging
operation (\ref{avinfty})
due to a generally nonvanishing first order term which arises from the expansion
of the
volume element $\sqrt{g^{(3)}} = \sqrt{g^B} ( 1 + h + O (h^2))$.
 
The most general representation of a symmetric $4\times4$ tensor $h_{\mu\nu}$
in terms of the complete basis of spherical harmonics is given by
\[
\ h_{00} = - 2 a^2 \sum_{n} A_n Q^{(0)}_n
\]
\be\label{exp}
\ h_{0i} = - a^2 \sum_{n} [ B^{(0)}_n Q^{(0)}_{n\;i} + B^{(1)}_n Q^{(1)}_{n\;i}]
\ee
\[
\ h_{ij} = 2 a^2 \sum_{n} [ H^{(0)}_{L n} g^{B}_{ij} Q^{(0)}_n
\                 +  H^{(0)}_{T n } Q^{(0)}_{n ij}
\                 +  H^{(1)}_{T n } Q^{(1)}_{n ij}
\                 +  H^{(2)}_{T n } Q^{(2)}_{n ij} ],
\]
where the coefficients
$A_n,B^{(0)}_n,B^{(1)}_n,H^{(0)}_{T n},H^{(1)}_{T n }$ and $H^{(2)}_{T n }$
are generally
dependent on the conformal time parameter $\bar{t}$.
Let $u^{\mu}$ be the four-velocity associated with the frame
in which the energy flux of the matter vanishes, then the
three-velocity $u^i / u^0$ associated with $u^{\mu}$, can
be expanded as
\be
\ u^i / u^0 = \sum_{n} [ v^{(0)}_n Q^{(0)}_{n\,i} + v^{(1)}_n Q^{(1)}_{n\,i} ],
\ee
where $Q^{(0)}_{n\,i} := - k_n^{-1} Q^{(0)}_{n |i}$,
and $u^0 = 1 / a (\bar{t}) $ to first order, due to the
normalization $u_{\mu}u^{\mu} = -1$.
 
A gauge transformation is defined as a change in the correspondence
between points $p$ in $S$, and points $\bar{p}$ in $\bar{S}$.
The most general first order gauge transformation is the result of the coordinate
transformation
\be\label{gauge}
\ \tilde{t} = t + \sum_n T_n Q^{(0)}_n (x^{\mu}),
\ee
\be\label{gauge2}
\ \tilde{x}^i = x^i +\sum_n (L^{(0)}_n Q^{(0) i}_n (x^{\mu}) + L^{(1)}_n
\  Q^{(1) i}_n (x^{\mu}) ),
\ee
in $S$, while the coordinates in $\bar{S}$ are fixed, and the correspondence
between points with the same coordinates in $S$ and
in $\bar{S}$ is kept fixed.
The coefficients $T_n$ and $L_n$ in expression (\ref{gauge}) and
(\ref{gauge2}) are arbitrary functions of the conformal time coordinate
$\bar{t}$. Notice that $T_n$ generates a change in the
correspondence  of the time coordinates in $S$ and
$\bar{S}$, while $L_n$ generates a change
in the correspondence between the spatial hypersurface coordinates on
$\Sigma$ and $\bar{\Sigma}$. The changes in the amplitudes of the metric
tensor are calculated in the case of scalar perturbations \cite{bardeen},
 \be\label{t1}
\ \tilde{A}_n = A_n - \dot{T}_n - \fr{\dot{a}}{a} T_n,
\ee
\be\label{t2}
\ \tilde{B}^{(0)}_n = {B}^{(0)}_n  + \dot{L}^{(0)}_n + k_n T_n,
\ee
\be\label{t3}
\ \tilde{H}^{(0)}_{L n} = H^{(0)}_{L n} - (k_n /3) L^{(0)}_n -
\ \fr{\dot{a}}{a} T_n,
\ee
\be\label{t4}
\ \tilde{H}^{(0)}_{T n} = H^{(0)}_{T n} + k_n L^{(0)}_n,
\ee
where a dot denotes conformal time differentiation.
For vector perturbations we find,
\be\label{v2}
\ \tilde{B}^{(1)}_n = {B}^{(1)}_n  + \dot{L}^{(1)}_n
\ \;\;\;,\;\;\;
\ \tilde{H}^{(1)}_{T n} = {H}^{(1)}_{T n} + k_n L^{(1)}_n,
\ee
while for tensor perturbations the trivial relation
$\tilde{H}^{(2)}_{T n} = H^{(2)}_{T n }$
holds for all $n$.
The matter velocity perturbation coefficients $v^{(0)}_n$ and $v^{(1)}_n$, with
respect to the coordinate frame, transform as,
\be\label{t10}
\ \tilde{v}^{(i)}_n =  v^{(i)}_n + \dot{L}^{(i)}_n,
\ee
where $i \in \{ 1,2 \}$.
Apart from the gauge freedom which is related to the mapping
between points in $S$ and $\bar{S}$,
there is a gauge freedom related to the choice of the
background scale factor $a (\bar{t})$.
A first order change in the choice of the background scale factor,
\be\label{bgt}
\ \tilde{a} (\bar{t}) =  a (\bar{t}) + D (\bar{t})
\ee
affects the spatially homogeneous mode of the
trace part of the spatial metric by a change,
\be\label{t7}
\ \tilde{H}^{(0)}_{L0} = {H}^{(0)}_{L0} + \fr{D}{a},
\ee
where ${H}^{(0)}_{L0}$ is the coefficient which multiplies
the spatially homogeneous trace mode in the expansion of
the metric (\ref{exp}), and
\be\label{snof}
\ {H}^{(0)}_{L0} = \fr{1}{3} \langle \langle \fr{\sqrt{g^{(3)}}}{\sqrt{g^B}} - 1
\ \rangle \rangle,
\ee
to first order.
 
The approach in this paper will be based on a specification of the
temporal part of the gauge (i.e., the correspondence  between the
time coordinates $t$ in $S$ and $\bar{t}$ in $\bar{S}$),
while maintaining covariance with respect to spatial gauge transformations
(i.e., the correspondence between the spatial coordinates
$x^i$ in $S$ and $\bar{x}^i$ in $\bar{S}$).
 
We stress that a fully gauge covariant approach is
preferred to an approach which is based on a (partially)
fixed gauge, since explicitly gauge dependent results
generally point out nonphysical features of the
calculation, including calculational mistakes.
However, the intricateness of a fully gauge covariant calculation at
second order makes such a calculation cumbersome (see e.g. \cite{bruni}),
and we will therefore follow an approach where the
temporal part of the gauge is fixed, while maintaining
spatial gauge covariance.
 
The temporal inhomogeneous part of the gauge is specified
by imposing conditions on the coefficients in the
expansion of the metric (\ref{exp}).
The extrinsic curvature tensor of the constant-$t$ hypersurfaces in $S$
is given by,
\[
\ K^i_j  =  \fr{1}{a} \sum_n \left[ \fr{\dot{a}}{a} +
\ ({\dot{H}}^{(0)}_{L n} - \fr{\dot{a}}{a} A_n + \fr{k}{3} B^{(0)}_n ) Q^{(0)}_n
\ \right] \delta^i_j
\]
\be\label{A0}
\  + \left[ {\dot{H}}^{(0)}_{T n} - k B^{(0)}_n \right] Q^{(0) i}_{n\;j}
\ + \left[ {\dot{H}}^{(1)}_{T n} - k B^{(1)}_n \right] Q^{(1) i}_{n\;j}
\ + {\dot{H}}^{(2)}_{T n} Q^{(2) i}_{n\,j}.
\ee
By requiring that the coefficients $A_n,B^{(0)}_n$ and $H^{(0)}_{L n }$ in the
expansion of the metric (\ref{exp}) satisfy
the condition
\be\label{g0}
\ \dot{H}^{(0)}_{L n} - \fr{\dot{a}}{a} A_n  + \fr{k}{3} B^{(0)}_n= 0 ,
\ee
for all $n$, we specify a gauge in which the hypersurfaces of constant time
$t$ in $S$
have spatially constant volume expansion $K =  3 \dot{a}/a^2$, as is clear by
contracting expression (\ref{A0}).
Condition (\ref{g0}) specifies more or less uniquely a collection
of spatial hypersurfaces in $S$ (see \cite{eardley}), but uniqueness
is not required in the calculation which follows, since, as
we will show in the following, our result for the average
expansion of an inhomogeneous universe does not in
relevant order depend on the choice of the
inhomogeneous temporal part of
the gauge.
 
Note that condition (\ref{g0}) does not constrain the choice
of the time coordinate in $S$, and the correspondence between the
time coordinates $t$ in $S$ and $\bar{t}$ in $\bar{S}$.
 We specify the time parameter $t$ in $S$, up to the freedom of
adding a constant, by imposing the requirement that the
homogeneous component of $A$ vanishes, i.e.,
\be\label{g1}
\ A_0 =0,
\ee
for all times $\bar{t}$.
 
The choice of gauge (\ref{g1}) implies that
the background time interval coincides with the averaged
proper time interval in $S$, as measured by observers which are
comoving with the spatial coordinates.
 
The gauge condition (\ref{g1}) can always be satisfied by performing a first
order homogeneous gauge transformation.
In order to clarify this statement, let us consider how equation (\ref{g1})
is affected by a homogeneous temporal gauge transformation.
According to expression (\ref{t1}), a homogeneous temporal gauge transformation
with $T=T_0$, induces a first order change in the metric perturbation
coefficient $A_0$,
\be\label{g00}
\  \tilde{A}_0 = A_0 - \dot{T}_0 - \fr{\dot{a}}{a} T_0.
\ee
The gauge condition (\ref{g1}) is satisfied by
performing a gauge transformation of the form (\ref{t1}), where
 
\be\label{cud}
\ T_0  =\fr{c}{ a (\bar{t})} +\fr{1}{a (\bar{t})}  \int^{\bar{t}}{\it d}\tau
\ a ( \tau ) A_0 (\tau) ,
\ee
and $c$ is a constant of integration.
The gauge condition (\ref{g1}) therefore determines the homogeneous
temporal part of the gauge, up to a constant of
integration $c$.
 According to the transformation law (\ref{t3}), the constant of integration $c$
in expression (\ref{cud}) affects the spatially homogeneous trace part of the
spatial metric.
This gauge freedom can be fixed by requiring that the homogeneous trace
perturbation of the spatial metric vanishes, i.e.,
\be\label{g2}
\ {H}^{(0)}_{L0} = 0,
\ee
for one time $\bar{t}_c$ and for a fixed choice of
the background scale factor $a (\bar{t})$ at
$\bar{t} = \bar{t}_c$.


Although we have completely specified the homogeneous temporal
part of the gauge by imposing the gauge conditions
(\ref{g1}) and (\ref{g2}), the homogeneous trace
perturbation of the spatial metric may still
differ from zero for times
$\bar{t} \neq \bar{t}_c$.
These perturbations are related to the freedom of choosing the background
scale factor $a (\bar{t})$ for times $\bar{t} \neq \bar{t}_c$, as is
clear from equation (\ref{bgt}).
 When we require that condition (\ref{g2}) holds at {\em all} times $\bar{t}$,
then it follows from equation (\ref{bgt}) and (\ref{t7}) that the choice of the
background scale factor $a (\bar{t})$ is fixed for all times
$\bar{t}  \in {\bf R}$.
 
Recall that in section (\ref{fff}) we derived the generic
linearized averaging operation for which unperturbed FLRW
is a stable fixed point.
It was shown that this linearized averaging operation which
works on the ten components of the metric tensor, reduces
to evaluating the spatial average of $\delta g^i_{\;i}$ and
$\delta g_{00}$.
By imposing the gauge conditions (\ref{g1}) and (\ref{g2}), we
specified a choice of background geometry by requiring that
the spatial averages of $\delta g^i_{\;i}$ and $\delta g_{00}$
both vanish.
For this choice of gauge, the averaged spacetime equals the background
spacetime, and the averaging problem reduces to solving the averaged
constraint equations for the background scale factor $a (\bar{t})$.
 
An explicit expression for the background scale factor $a (\bar{t})$
in the gauge fixed by condition (\ref{g2})
is obtained by substituting the expression for
the background metric (\ref{bg}) and expression (\ref{exp}) for the
perturbed metric, into expression (\ref{snof}).
To first order we find,
\be\label{at}
\ a^{3} (\bar{t}) =  \langle \langle \fr{\sqrt{g^{(3)}}}{\sqrt{\eta}}
\ \rangle \rangle,
\ee
where $g^{(3)} = \det(g_{ij})$ and $\eta = \det(\eta_{ij})$.
 
Recall that condition (\ref{g0}) fixes the inhomogeneous temporal part
of the gauge, and the collection of spatial hypersurfaces on which
the spatial average is evaluated.
Since physical results must be gauge invariant to relevant order,
one may question whether the freedom of choosing a family
of hypersurfaces affects the result for the
scale factor (\ref{at}).
It follows from the orthonormality relation (\ref{nrm}) and the
transformation property (\ref{t3}) that the background
scale factor (\ref{at}) is invariant to first
order under inhomogeneous temporal gauge
transformations.
However, at order $\epsilon^2$ inhomogeneous metric perturbations do
contribute to the background scale factor (\ref{at}), and the
gauge invariance of the scale factor $a (\bar{t})$
therefore breaks down at order $\epsilon^2$.
Consistent with this limitation we will neglect terms of order $\epsilon^2$
in our calculation, while retaining terms of order
$\epsilon$ and $\epsilon^2 / \kappa^2$.

Summarizing the content of this subsection, we completely specified
the temporal and the spatially homogeneous part of the gauge, and
the choice of the background,  by imposing the gauge
conditions (\ref{g0}) (\ref{g1}) and (\ref{g2}) on the metric
coefficients $A_n,B^{(0)}_n$ and $H^{(0)}_{L n }$.
 
\section{Averaging the constraint equations}\label{ace}
 
The classical constraint equations on a
hypersurface $\Sigma$ are given by
\be\label{constr1}
\ R^{(3)} + K^2 - K_{i j} K^{i j} = 16 \pi G \rho + 2 \Lambda,
\ee
\be\label{constr504}
\ K^{j}_{i;j} - K_{;i} = 8 \pi G J_i,
\ee
where $;$ denotes the covariant derivative with respect to
$g_{i j}$, and $R^{(3)}$
is the Ricci scalar associated with the induced
metric $g_{i j}$, and
\be\label{rho}
\ \rho = T^{\mu \nu}  n_{\mu} n_{\nu} \;,\;\; J_i =
\ - T^{\mu \nu} h_{i \mu} n_{\nu},
\ee
where $n_{\mu}$ denotes the future directed unit
vector normal to ${\Sigma}$, and
$h_{\mu\nu} := g_{\mu\nu} + n_{\mu} n_{\nu}$.
 
In the constant-$K$ gauge, defined by conditions (\ref{g0}),
(\ref{g1}) and (\ref{g2}),
the constraint equation (\ref{constr1}) takes
the form \be\label{constrz}
\ \fr{\dot{a}^2}{a^4}  = \fr{8\pi}{3} G \rho
\ - \fr{1}{6} R^{(3)}
\ + \fr{1}{6} \hat{K}_{i j} \hat{K}^{i j}  + \fr{1}{3} \Lambda,
\ee
where $\hat{K}^{i j} := K^{ij} - \fr{1}{3} g^{ij} K$ is the traceless part
of the extrinsic curvature tensor.

In principle one could solve the constraint equation
(\ref{constrz}) for the time dependence of the
scale factor $a (\bar{t})$, while taking into account all
linear and higher order contributions to the
right-hand side of equation (\ref{constrz}).
However, this approach is unnecessarily complicated, since all
terms which do not have constant values on $\Sigma$ must cancel
on the right-hand side of equation (\ref{constrz}),
since the left-hand side of equation (\ref{constrz}) is constant
on $\Sigma$.
For the sake of {\em calculational} convenience, we will
take the spatial average at the right-hand side
of the constraint equation (\ref{constrz}), without
changing any physical aspects of the
constraint equation;
\be\label{constrxx}
\ \fr{\dot{a}^2}{a^4}  = \fr{8\pi}{3} G \langle \rho  \rangle
\  - \fr{1}{6}  \langle R^{(3)} \rangle
\ + \fr{1}{6}  \langle \hat{K}_{i j} \hat{K}^{i j} \rangle + \fr{1}{3}\Lambda.
\ee
In order to solve equation (\ref{constrxx}) for the scale factor $a (t)$,
we need to evaluate the spatial average of the 3-curvature $R^{(3)}$,
and the energy density $\rho$, and the square of the
traceless part of the extrinsic curvature tensor
$\hat{K}_{i j} \hat{K}^{i j}$.
We will calculate these quantities in the following three subsections.
 
\subsection{The averaged spatial curvature}\label{asc}
 
The spatial curvature perturbation $\delta R^{(3)}$ can be
expanded in terms of the 3-metric perturbation $h_{ij}$
(see e.g. \cite{tHooft}),
 
\be\label{eqd}
\  R^{(3)} = \fr{6 {\bf k}}{a^2}  + \delta R^{(3)} ,
\ee
where
\be\label{cup}
\ \delta R^{(3)} =   \delta R   +
\  \delta^2 R + O (h^3),
\ee
and
\be\label{e0}
\ \delta R =    h^{|k}_{|k} - h^{k|i}_{\;i|k},
\ee
\be\label{e1}
\ \delta^{2} R = -  \fr{1}{4} h^{ij} h^{\; |q}_{ij|q}
\ + \fr{1}{2} h^{ij} h^{\;|q}_{qi|j}
\ - \fr{1}{4} h h^{|l}_{|l} + \mb{td.},
\ee
where td. stands for terms which are total derivatives.
Let us now evaluate the contributions to the averaged curvature perturbation
$ \langle R^{(3)} \rangle$ for scalar, vector, and tensor modes in the expansion
of $h_{ij}$.
 
\subsubsection{Scalar perturbations}\label{scp}
 It follows from expression (\ref{cup}) that the lowest order
contribution to the spatial curvature perturbation is given by
$g^{B ij} \delta R_{ij}$, which is order $\epsilon / \kappa^2$,
but the spatial average of this contribution vanishes to
order $\epsilon / \kappa^2$, due to the orthogonality
relations (\ref{nrm}).
The linear curvature perturbation $g^{B ij} \delta R_{ij}$ does however
contribute
to the averaged 3-curvature perturbation by a term of order
$\epsilon^2 / \kappa^2$, i.e.,
\be\label{eqd2}
\  \langle  \delta R  \rangle
\ =  \fr{12}{a^2} \sum_n  ( k_n^2 - 3 {\bf k})  H^{(0)}_{L n } ( H^{(0)}_{L n } +
 \fr{1}{3}
\  H^{(0)}_{T n })   \ + O (\epsilon^3 / \kappa^2),
\ee
where we made use of the expansion of the volume element
$\sqrt{g^{(3)}} = \sqrt{g^B} ( 1 + h + O (h^2))$, and the
definition of the spatial average (\ref{avinfty}).

The quadratic term $\delta^2 R$ in the expansion of
the 3-curvature perturbation (\ref{cup}) contributes
to the averaged 3-curvature perturbation by a term
\be\label{eqd22}
\  \langle   \delta^2 R \rangle ^{(0)}_{ \infty}
\ = -  \fr{1}{a^2}  \sum_n  ( k_n^2 - 3 {\bf k})
\ (  10 H^{(0)\, 2}_{L n} - \fr{2}{9}  H^{(0)\,2}_{T n}
\  + \fr{8}{3} H^{(0)}_{T n } H^{(0)}_{L n} )  + O (\epsilon^3 / \kappa^2),
\ee
where we used the computer algebra package MAPLE to derive this expression.
Combining expression (\ref{eqd2}) and (\ref{eqd22}), we find an
expression for the scalar contribution to the
spatial curvature perturbation,
\be\label{fequ}
\  \langle \delta R^{(3)} \rangle ^{(0)}  =  \fr{2}{a^2}  \sum_n (k_n^2 -3{\bf k}
)
\phi^2_{h n}  + O (\epsilon^3 / \kappa^2),
\ee
where
\be\label{phih}
\ \phi_{h n} := H^{(0)}_{L n } + \fr{1}{3} H^{(0)}_{T n }
\ee
is the gauge invariant amplitude which measures the distortion of the
intrinsic geometry of the constant-$K$ hypersurfaces.
Using the expansion (\ref{cup}) for the spatial curvature perturbation,
and the definition (\ref{phih}) of $\phi_{h n}$, one finds that
$\phi_{h n}$ is related to the {\em first order} spatial
curvature perturbation by,
\be\label{R1}
\ \delta R  = \fr{4}{a^2} \sum_n (k_n^2 - 3 {\bf k})  \phi_{h n} Q^{(0)}_n .
\ee
By substituting expression (\ref{R1}) into the
constraint equation (\ref{constr1}), we obtain a simple expression
for $\phi_{h n}$ in terms of the first order energy perturbation,
\be\label{constr111}
\  \phi_{h n} = \fr{ 4 \pi a^{2}}{ (k_n^2 - 3{\bf k})} G \bar{\rho}
\ \epsilon_{h n}  + O (\epsilon^2/ \kappa^2),
\ee
for all $n$, where $\epsilon_{h n}$ is defined as the
density contrast
in the constant-$K$ gauge,
\be\label{epsilonh}
\ \epsilon_{h n} = \delta \rho  (k_n) / \bar{\rho} ,
\ee
and $\bar{\rho}$ denotes the background energy density.
 
 We would like to express the scalar contribution to the averaged curvature
perturbation in terms of observable quantities.
Since the averaged curvature perturbation (\ref{fequ}) is quadratic in $\phi_{h n}$,
we may use the constraint equation (\ref{constr111}) to {\em first order}
to determine $\phi_{h n}$ in terms of the fractional
energy perturbation $\epsilon_{h n}$. We obtain,
\be\label{xequ}
\  \langle \delta R^{(3)} \rangle^{(0)}  = 32 (\pi a G \bar{\rho})^2
\     \sum_n  \fr{\epsilon^2_{h} (q_n) }{ (k_n^2 - 3{\bf k})}
\  + O (\epsilon^3/\kappa^2),
\ee
where the sum (or integral when $\Sigma$ is open) is taken over all
possible values $n$.
Expression (\ref{xequ}) takes an especially simple form when
expressed in terms of the power spectrum $P (k)$,
which allows the representation
\be\label{power}
\ P_h (k ) =   \sum_n \fr{ \epsilon_{h }^2 ( q_n ) }{4\pi q_n^2 }
\  \delta ( k - |q_n| ),
\ee
where the subscript $h$ refers to the constant-$K$ gauge
(see e.g. \cite{peebles} or \cite{coles} for more on
power spectra).
Combining expressions (\ref{xequ}) and (\ref{power}) yields,
\be\label{cvhx}
\  \langle \delta R^{(3)} \rangle^{(0)}  =  32 ( \pi^2  G {\bar{\rho}} )^2
\   J_2 + O (\epsilon^2,\epsilon^3/\kappa^2),
\ee
where
\be\label{j2}
\  J_2 := 4 \pi a^2 \int_0^{\infty} {\it d} k P_h (k)
\ee
is an observable quantity known as the second
moment of the power spectrum, and by absorbing a factor 
$a^2$ in the definition of $J_2$ we restored
physical units of length square.

\subsubsection{Vector perturbations}

Using the definition of the vector harmonics (\ref{eee1}), and
the orthogonality relations (\ref{nrm}), we find that
vector perturbations do not contribute to the
spatial curvature perturbation (\ref{cup}).
This result may be expected, since it follows from expression
(\ref{v2}) that one can always choose a gauge in
which there are no vector perturbations of the
spatial metric, and the vector contribution
to the averaged spatial curvature perturbation
(\ref{cup}) must therefore vanish
in any gauge, due to gauge invariance of the
averaged spatial curvature perturbation.
 
\subsubsection{Tensor perturbations}\label{tp}
 
Using equations (\ref{nrm}), and expression (\ref{cup}) for the
second order expansion of the spatial curvature, it follows immediately that,
\be\label{st}
\  \langle \delta R^{(3)} \rangle^{(2)}  =  \fr{1}{a^2} \sum_n  k_n^2
\  H^{(2)\,2}_{T n},
\ee
while the tensor contribution to the term
$\langle \hat{K}_{i j} \hat{K}^{i j} \rangle$
in the averaged constraint equation (\ref{constrz}) follows immediately
from the expression for the extrinsic curvature tensor (\ref{A0}).
Although the tensor contribution to the averaged constraint equations
is easily calculated in terms of the coefficients $H^{(2)}_{T n }$,
the magnitude of this term has not yet been determined quantitatively
by the observation of gravitational waves.
 
\subsection{Averaged energy density}\label{aed}

In this subsection we will calculate the averaged energy
density $ \langle \rho \rangle$.
In order to calculate the lowest order nontrivial contribution to the
averaged energy perturbation, we will adopt the assumption in this
subsection that the matter in the universe at late times after
decoupling can be effectively described by the energy momentum
tensor density for a pressureless perfect fluid, i.e.,
\be\label{perf}
\ T^{\mu\nu} = \rho_0 u^{\mu} u^{\nu},
\ee
where $u^{\mu}$ is the four-velocity of the fluid, and $ \rho_0$
is the energy density in the rest-frame of the fluid.
The equations of motion for the fluid read,
\be\label{emo}
\ \nabla_{\mu} T^{\mu\nu} = 0,
\ee
which implies
\be\label{rho_0}
\  \p_{\mu} \left( \sqrt{-g} \rho_0 u^{\mu}\right)  = 0,
\ee
where we used that
$\nabla_{\mu}= (\sqrt{-g})^{-\fr{1}{2}} \p_{\mu} \sqrt{-g}$, and
$ u^{\nu} \nabla_{\nu} u^{\mu} =0$
for a pressureless fluid.
By using the spatial gauge freedom (\ref{gauge2}) we may set
$B^{(0)}_n = B^{(1)}_n =0$, such that
$\sqrt{-g} = \sqrt{-g_{00}} \sqrt{g^{(3)}}$, and the equation of
motion (\ref{rho_0}) takes the form
\be\label{kdl}
\  \p_{\mu} \left( \sqrt{-g_{00}}  \sqrt{g^{(3)}} \rho_0 u^{\mu} \right) =0,
\ee
while in this gauge $u^i / u^0$ equals the matter 3-velocity with respect to the
normals to the constant-$K$ hypersurfaces.
The velocity four-vector $u^{\mu}$ can be written in the form
\be\label{kkh}
\ u^{\mu} = \left[ \fr{1 + v^2_{h} }{-g_{00}} \right]^{\fr{1}{2}}
\ \delta_0^{\mu} + u^i \delta^{\mu}_i,
\ee
where $v^2_h :=  g_{ij} u^i u^j$ equals to first order the square
of the velocity three-vector $u^i / u^0$, and
we used that $u^{\mu} u_{\mu} =-1$.
By substituting expression (\ref{kkh}) into the equation of motion (\ref{kdl}),
we find
\be\label{clasc}
\ \fr{\p}{\p t} \left[ (1 + \fr{1}{2}v^2_h) \sqrt{g^{(3)}} \rho_0 \right]
\ + \fr{\p}{\p x^i} \left[ \sqrt{-g_{00}} \sqrt{g^{(3)}} \rho_0 u^i \right]  =0,
\ee
to first order.
Using equation (\ref{clasc}) and the definition of the spatial average
(\ref{avinfty}), we obtain
\[
\ \lim_{\ell \rar \infty} \fr{\p}{\p t}
\   \langle (1 + \fr{1}{2}v^2_h )  \rho_0 \rangle (\ell)  =
\]
\[
\ - \lim_{\ell \rar \infty}  ( \fr{\p}{\p t} \ln  \tilde{N} (x,\ell))
\   \langle (1 + \fr{1}{2}v^2_h )  \rho_0 \rangle (\ell)
\]
\be\label{ppy}
\ - \lim_{\ell \rar \infty} N^{-1} (\ell)
\ \int_{\bar{\Sigma}} {\it d} x^{\pr}
\ \fr{\p}{\p x^i}  \sqrt{-g_{00}}  \sqrt{g^{(3)}}  \rho_0 u^i
\  \theta ( \ell - \Delta s (x,x^{\pr}) ) ,
\ee
where $\tilde{N} (\ell)$ denotes the dimensionless quotient of $N (\ell)$,
and a constant with the dimension of a 3-volume.
The second term on the right-hand side of equation (\ref{ppy}) vanishes
due to Gau\ss's theorem. Combining the remaining terms in equation
(\ref{ppy}) yields,
\be\label{qxq}
\  \lim_{\ell \rar \infty} \fr{\p}{\p t} \mb{ln}
\ \langle (1 + \fr{1}{2}v^2_h ) \rho_0
\ \rangle (\ell)  \ = - \lim_{\ell \rar \infty} \fr{\p}{\p t} \mb{ln}
\   \tilde{N} (x,\ell) .
\ee
By integrating equation (\ref{qxq}), it follows that
\be\label{bbv}
\ \langle ( 1 + \fr{1}{2}v^2_h )  \rho_0 \rangle (t)  \propto
\ \fr{1}{\tilde{N} (x,\ell)}
\ \propto \fr{a^3 (t_0)}{a^3 (t)} ,
\ee
where we used the gauge condition (\ref{g2}).
Formula (\ref{bbv}) shows that the rest-frame energy density $\rho_0$, when
integrated over a spatial volume element on $\Sigma (t)$ which is comoving
with the matter flow,
is not conserved for a pressureless fluid, while
$\rho_0 (1 + \fr{1}{2} v^2_h )$ is
conserved to first order.
 
The spatial average of the energy perturbation $\delta \rho$ is obtained
by expanding equation (\ref{perf}) for $T^{00}$ to first order, where
we use (\ref{kkh}) and the gauge condition (\ref{g1}). We find
\be\label{xxx}
\ \langle \rho  \rangle (t) = \langle ( 1 +  v^2_h ) \rho_0 \rangle (t),
\ee
which combines with equation (\ref{bbv}),
\be\label{fgh}
\ \langle \rho  \rangle (t) = \bar{\rho} (t)
\ + \langle \fr{1}{2} \bar{\rho} v^2_h  \rangle (t),
\ee
to first order, where we used that   
$\langle \rho_0  \rangle (t_0)$ equals $\bar{\rho} (t_0)$ 
when perturbations vanish at time $t_0$.
Indeed, the lowest order contribution to the averaged energy
density (\ref{xxx}), is given by the sum of the averaged
rest-mass of the fluid, and the
(nonrelativistic) kinetic energy of the fluid.
Since $v^2_h$ is of the order of $\epsilon$, the lowest order
correction to the averaged energy perturbation is typically
small in the observed universe, but nevertheless
significant in the sense of the ambiguity which is related
to the freedom of choosing a gauge
and an averaging operation (see section \ref{gp}).
 
It is interesting to note that there exists a simple relation
due to Irvine and Layzer (see e.g., \cite{peebles}) which relates
$W:= 2 \pi G \bar{\rho} J_2$, where $J_2$ is defined
by equation (\ref{j2}), and the energy due to the peculiar velocity
$L := \langle \fr{1}{2} \bar{\rho} v^2_h  \rangle$.
For a pressureless fluid and nonrelativistic motions, it can
be shown that
$ \fr{\p}{\p t} (a W - a L )  =  L \dot{a}$,
which, assuming that the universe departs from small values
of $J_2$ and $L$ and relaxes to a nearly time independent
bound state at late times, implies the Newtonian virial
theorem $L=W/2$.

Note that our result differs from a result derived by Futamase
(see \cite{futamase} and \cite{futamase2}), where one finds
a peculiar velocity contribution to the averaged energy density
which is exactly twice as large as our result (\ref{xxx}).
This result seems to be based on the erroneous assumption that
the integral of rest-frame energy density over a spacelike
hypersurface is time independent (this is only true in a gauge where
$v^i - B^i $ vanishes). In this case, equation (\ref{perf}) yields
an averaged energy perturbation which is twice the result (\ref{xxx}).
However, this result violates continuity of the scale factor
at the right-hand side of equation (\ref{constrxx}) when
rest-mass is instantaneously and homogeneously converted
into kinetic energy or vice versa.
 
\subsection{The squared shear contribution}\label{sqs}
 
In this subsection we will evaluate the contribution of the term
\be\label{sqsr}
\ \langle  \hat{K}_{ij} \hat{K}^{ij} \rangle,
\ee
in the averaged constraint equations (\ref{constrxx}), for
scalar and vector perturbations.
The scalar and vector part of $\hat{K}_{ij}$ are coupled to the matter
current by the constraint equation (\ref{constr504}),
which takes the form,
\be\label{constr504x}
\ \hat{K}^{j}_{i;j}  = 8 \pi G J_i,
\ee
when evaluated in the constant-$K$ gauge.
The matter current $J_i$ is defined by expression (\ref{rho}),
and can be expanded as
\be\label{exj}
\ J_i = (\bar{\rho} + \bar{P}) \sum_n \left[ v^{(0)}_{h n} Q^{(0)}_{n i}
\  + v^{(1)}_{h n} Q^{(1)}_{n i} \right] ,
\ee
to first order, where $v^{(0)}_{h n}$ and $v^{(1)}_{h n}$ denote the scalar and
vector components of the velocity three-vector of the matter with respect
to the normals to the constant-$K$
hypersurfaces, $Q^{(0)}_{n i} := - k^{-1} Q^{(0)}_{n | i}$,
and $\bar{\rho}$ and $\bar{P}$ denote the background
energy and pressure density.
 
By substituting the traceless part of the extrinsic curvature tensor
(\ref{A0}), and the expansion (\ref{exj})
for $J_i$, into the constraint equation (\ref{constr504x}), we obtain
\be\label{bdg}
\  \fr{2}{3} (k_n - 3 {\bf k} /k_n) [ \dot{H}^{(0)}_{T n} - k_n B^{(0)}_n ] =
\  a G (\bar{\rho} + \bar{P})
\ v^{(0)}_{h n}
\ee
for scalar perturbations, and
\be\label{gsd}
\  \fr{1}{2} (k_n - 2 {\bf k} /k_n) [ \dot{H}^{(1)}_{T n} - k_n B^{(1)}_n ] =
\ a G (\bar{\rho} + \bar{P})
\ v^{(1)}_{h n}
\ee
for vector perturbations.
Expressions (\ref{bdg}) and (\ref{gsd}) yield expressions
for the scalar and vector traceless part of the extrinsic curvature
tensor (\ref{A0}),
in terms of the matter velocity, which can be used to
evaluate the scalar and vector contribution to
expression (\ref{sqsr}).
For scalar perturbations we find,
\be\label{cn1}
\ \langle  \hat{K}^{(0)}_{ij} \hat{K}^{(0) ij} \rangle
\ = \fr{3}{2} a^2 G^2 (\bar{\rho} + \bar{P})^2
\  \sum_n  \fr{v_{h n}^{(0) 2}}{(k_n - 3{\bf k}/k_n)^2},
\ee
and for vector perturbations,
\be\label{cn2}
\ \langle  \hat{K}^{(1)}_{ij} \hat{K}^{(1) ij} \rangle
\ = 2 a^2 G^2  (\bar{\rho} + \bar{P})^2
\  \sum_n  \fr{v_{h n}^{(1) 2}}{(k_n - 2{\bf k} /k_n)^2}.
\ee
The coupling between the matter current and the shear of the normals
to the constant-$K$ hypersurfaces, can be interpreted as the
`frame dragging' effect which occurs in the presence of
moving matter (e.g., as in the region around a
rotating black hole).
It follows from expressions (\ref{cn1}) and (\ref{cn2}), taking into account
the normalizations of the scalar and vector modes (see expression (\ref{q3})),
that the
matter current and $\hat{K}_{ij}$ couple with different
strength for scalar and vector perturbations.
Furthermore, the strength of the coupling vanishes proportional to
$k_n^{-1}$ when $k_n \rar \infty$.
Since $v_{h n}^2 = O (\epsilon)$, when velocity perturbations are
generated by density perturbations at late times, it follows that
expressions (\ref{cn1}) and (\ref{cn2}) contribute to the
averaged constraint equations (\ref{constrxx}) by a term
of order $\epsilon \kappa^2$, which is negligible compared
to the leading order kinetic energy contribution
discussed in section (\ref{aed}) for perturbations
at length scales much smaller than the Hubble radius.
 
However, for perturbations at arbitrary large length scales,
the strength of the coupling grows proportional
to $\delta^{-1}$ when $\delta \downarrow 0$, where
$\delta := k_n^2 - 3 {\bf k}$ for scalar
perturbations and
$\delta := k_n^2 - 2 {\bf k}$ for vector
perturbations.
Note that since $k_n$ must be real for bounded solutions, the limit
$\delta \downarrow 0$ does not exist when ${\bf k} < 0$, and the
limit $\delta \downarrow 0$ does not exist when ${\bf k} >0$
since $k_n$ takes only discrete values
in this case.
 
Note that the divergent coupling between the metric and the matter velocity
for $\delta \downarrow 0$ and ${\bf k} = 0$, is unrelated to the dynamics of the
matter and metric at small scales and late times, since perturbations
for which $\delta \ll 1$ are typically larger than the Hubble radius,
and must have a primordial origin.
 
A natural question which arises is whether the divergence in equations
(\ref{cn1}) and (\ref{cn2}) for $\delta \downarrow 0$ can be purely
attributed to a large warping of the constant-$K$ hypersurfaces, which
can be removed by choosing another gauge.
Indeed, it follows from expressions (\ref{A0}) and (\ref{t2}) that the
scalar part of
$\hat{K}^i_{j}$ can be set equal to zero, by a temporal gauge transformation
with $T = k^{-2} [ \dot{H}^{(0)}_{T n} - k B^{(0)}_n ]$, but according to
expressions (\ref{R1}), (\ref{t3}) and (\ref{t4}), the {\em intrinsic}
spatial curvature does diverge when $\delta \downarrow 0$ in this gauge.
Furthermore, due to expression (\ref{v2}), the vector part of
$\hat{K}^i_{j}$ is gauge invariant, and the divergence in equation
(\ref{cn2}) is therefore independent of the choice of time-slicing.
 From the point of view of the matter, the most natural choice
of gauge is a comoving time-orthogonal gauge, which is defined by the condition
that the spatial
coordinates are comoving with the normals to the constant-$t$
hypersurfaces (i.e., $B^{(0)}= B^{(1)} = 0$), and the scalar part of the matter
velocity with respect to the normals to the constant-$t$
hypersurfaces vanishes (i.e., $v^{(0)} - B^{(0)} = v^{(0)} = 0)$.
According to expressions (\ref{t2}) and (\ref{t10}), a gauge
transformation from the constant-$K$ gauge to a comoving time-orthogonal
gauge is generated by $T = k^{-1} v^{(0)}_{h n}$.
In this gauge, the scalar part of the shear of the
{\em matter} coincides with the scalar part of
$\hat{K}^i_{j}$.
By transforming equation (\ref{bdg}) from the constant-$K$ gauge to a
comoving gauge, we find that the infra-red divergence of the scalar part
of $\hat{K}^i_{j}$ has the same strength in both gauges,
and its presence is therefore related
to the presence of shearing matter.
At first sight, a divergence of the shear of the matter for $\delta \downarrow 0$
seems to be inconsistent with the smallness of the velocity
perturbations which are the source of the
metric perturbations.
There is no real inconsistency however, since the matter velocity
perturbation is gauge dependent, and it might therefore
be anomalously small in the constant-$K$ gauge,
without being in conflict with large
matter shear perturbations.
These observations show that the divergence in equations
(\ref{cn1}) and (\ref{cn2}) is of a physical nature.

The absence of FLRW solutions of the constraint equations (\ref{constr504x})
when homogeneous vector perturbations of the matter velocity are present,
might seem peculiar, since solutions of the Einstein
equations correspond to stable points of the action.
At this point we should recall that we have limited our scope
to FLRW background spacetimes, which are by definition
spatially homogeneous and isotropic.
In the presence of homogeneous matter velocity perturbations,
our spacetime is no longer isotropic in the averaged sense, and
there is no FLRW background solution which is everywhere close
to our perturbed spacetime.
A satisfactory description of homogeneous velocity perturbations 
about FLRW, requires the inclusion of background 
solutions which are homogeneous but not necessarily 
isotropic, and which include FLRW as a special case. 
These solutions are given by the Bianchi models
of type V and VII$_h$, which include FLRW with ${\bf k} = -1$
as a special case, and type VII$_0$ which includes 
FLRW with ${\bf k} = 0$ (see e.g., \cite{hawking-collins} --
\cite{ellis-wainwright}).

\subsection{The averaged expansion}\label{rdn}
 
By substituting the expressions for the averaged curvature perturbation
and the averaged energy density, which where derived in the previous
subsections \ref{asc}, \ref{aed} and \ref{sqs},
into the averaged constraint equation
(\ref{constrxx}), we obtain,
\be\label{constrj}
\ \fr{\dot{a}^2 }{a^4}  =
\  \fr{8\pi}{3} G  \bar{\rho}
\  - \fr{{\bf k}}{a^2} +  \fr{1}{3} \Lambda
 +  \fr{8\pi G}{6} \langle \bar{\rho}  v^2_h \rangle
\ - \fr{32 \pi^2}{3} ( G{\bar{\rho}} )^2  J_2
\ee
\[
\ + \mb{g.w.}
\ + O (\epsilon^2,\epsilon \kappa^2,\epsilon^3/\kappa^2),
\]
where $J_2$ is defined by equation (\ref{j2}), and the term g.w.
denotes the contribution due to gravitational waves
(see subsection \ref{tp}).
We see that the averaged constraint equation (\ref{constrj})
takes the form of the standard Friedmann equation, plus
a contribution due to the peculiar velocity of the matter,
and a contribution due to the averaging of scalar and
tensor metric perturbations.
Let us now determine the magnitudes of the different contributions
on the right-hand side of equation (\ref{constrj}) by means of
the observational values for $\bar{\rho}$ and $J_2$.
Estimates from the Lick
and CfA catalogs \cite{lick} \cite{C&B} value
$J_2 \approx 200 h^{-2}$ Mpc$^2$, and
$\bar{\rho} \approx  1.88 \times 10^{-29} h^2 \Omega$ g cm$^{-3}$,
where $h$ is a dimensionless factor which expresses
the uncertainty in the value of the Hubble parameter
$H_0 = 100 h$ km s$^{-1}$ Mpc$^{-1}$, and $h$ is believed to be
between $0.5$ and $0.85$.
Inserting these values in the different terms on the right-hand side
of equation (\ref{constrj}), one finds,
\be\label{m22}
\ \fr{8 \pi G}{3} \bar{\rho} = 1.14 \times 10^{-35} h^2 \Omega \mb{s}^{-2},
\ee
\be\label{m33}
\ \fr{32 \pi^2}{3} ( G{\bar{\rho}} )^2  J_2 \sim
\ 1.0 \times 10^{-39} h^2 \Omega^2 \mb{s}^{-2},
\ee
and
\be\label{m11}
\ \fr{8\pi G}{6} \langle  \bar{\rho}  v^2_h \rangle \sim 1.3 \times 10^{-40} h
\ \Omega \mb{s}^{-2},
\ee
where we used the relation $v \sim (3 \pi G \bar{\rho} J_2 )^{1/2}$
(see section \ref{aed}).
According to equations (\ref{m22})-- (\ref{m11}), and the
constraint equation (\ref{constrj}), the matter induced metric
inhomogeneities act as a very small negative correction to the averaged
energy density, equal to about 1.0 $\times \Omega$ part in $10^4$,
while the back-reaction due to the peculiar velocity of the
matter acts as a positive
correction to the averaged energy density, equal to
about $1.2$ parts in 10$^5$.
The small negative correction to the averaged energy density
leads to a slight overestimation of the age of the universe
$t_0 = \fr{2}{3} H_0^{-1}$ assuming that $\Omega =1$, equal to about
$5$ parts in 10$^5$.

\section{Comparison with previous work}
 
The work on this paper started as a correction of the derivation
by Futamase in \cite{futamase} \cite{futamase2} on the points of
the treatment of the gauge freedom (see section \ref{gp})
and the choice of the averaging operation
(see section \ref{fff}). The paper was also inspired as an attempt to
address the fundamental ambiguity which enters the calculation
of any averaged metric through the freedom of choosing an
averaging operation.
 
In a recent independent paper by Russ {\it et al} \cite{russ},
the back-reaction due to density perturbations was calculated
by using the relativistic Zel'dovich approximation \cite{zeldovich}
in a comoving gauge.
The expression derived by Russ {\it et al} for the back-reaction
due to matter density perturbations agree in sign, but is roughly
an order of magnitude larger than the result derived in this paper.
Furthermore, a possible effect due to vorticity of the matter was 
ignored in that paper.
It should be noted that direct comparison between the results
by Russ {\it et al} and the results derived in this paper, is
nontrivial due to the fact that the gauges used in the two
papers are not related by a first order gauge transformation.
Namely, a gauge transformation from the constant-$K$ gauge
to the comoving synchronous gauge requires
$\dot{L}_n = - v_{h n}$ due to equation
(\ref{t10}), and $v_{h n} = O (\epsilon^{1/2})$ since
$v_{h n}^2 = O (\epsilon)$ when velocity perturbations
are generated by density perturbations at late times.
By working in a constant-K gauge, we avoided the problem of a breakdown
of the perturbative expansion which occurs in the comoving gauge
(namely, since metric and matter density perturbations are of the same
magnitude in a comoving gauge, metric perturbations get typically
large at late times, even though the perturbations in the intrinsic
geometry are generally small in the observed universe).
 
Finally, we mention the paper by Buchert and Ehlers \cite{ehlers},
where one integrates the Raychaudhuri equation over a spatial
hypersurface in a Newtonian background, and a globally vanishing
correction to the averaged expansion was found.
Although the Raychaudhuri equation is also valid in GR, the Newtonian
approximation enters the calculation where the correction to the
averaged expansion is expressed in terms of a boundary term,
which accounts for the difference between the Newtonian result
and the nontrivial correction (\ref{m22}) to the averaged
energy density derived in this paper.

\section{Conclusions}\label{avop}
 
We derived the generic linearized averaging operation for metrics
starting from the requirement that unperturbed FLRW is a stable
fixed point of the averaging operation.
By a gauge invariant approach, we eliminated
unphysical degrees of freedom in our problem,
and we explicified the fundamental ambiguities
which are related to the freedom of fitting the
averaged spacetime to the inhomogeneous spacetime.
The leading order nontrivial corrections to the
standard Friedmann equation are expressed in terms
of the power spectrum of the matter, and the effect
is calculated quantitatively by means of the
observational data.
The dominant correction to the averaged expansion is caused by
the back reaction of matter density perturbations, and leads
to a slower expansion rate and an overestimation of the age
of the universe by approximately $5$ parts in $10^5$.
The back-reaction of velocity perturbations, including vortical
motion of the matter, appears to be negligible at small
length scales. However, it was shown that the back-reaction of 
velocity perturbations can be 
significant in the large wavelength limit.
 
\section{Acknowledgements}
 
I would like to thank especially George F. R. Ellis
for helpful comments.
Thanks also go to Henk van Elst, and to Mauro
Carfora for reading the manuscript and inviting
me to SISSA (Italy), where part of the work
was done.
The research was supported with funds from
UCT (Cape Town).

\section{Appendix A}

In this appendix, we will discuss the relation between the
volume element in the hypersurface integral, and
gauge invariance at second and higher order in the
expansion parameter of the gauge transformation.

Let $\phi$ be a one parameter group of diffeomorphisms
$\phi : {\bf R} \times \Sigma \rightarrow \Sigma$, which
is defined by the condition that $\phi_{\lambda = 0 }$ is the
identity, and the curves $\phi_{\lambda} (p)$ are integral
curves of a vector field $\xi$ in $\Sigma$
(see e.g. \cite{wald} and \cite{bruni} for the mathematical
details which are involved).
A gauge is specified by choosing a mapping between points $p$ in $\Sigma$,
and points $\bar{p}$ in $\bar{\Sigma}$. Assuming that a choice of gauge
has been made, then a one parameter {\em group} of gauge choices is obtained
by mapping the points $\phi_{\lambda} (p)$ in $\Sigma$ to
points $\bar{p}$ in $\bar{\Sigma}$, for all $\lambda \in {\bf R}$
(the more generic case of a one parameter {\em family} of mappings
of points in the background and the perturbed spacetime, is
discussed in \cite{bruni}, but there is no need to introduce this
complication in the derivation which follows).
 
Let us consider a scalar function $q (x)$, which lives in $\Sigma$
(such that its value in a point $p$ in $\Sigma$ is fixed, while its
value in a point $\bar{p}$ in $\bar{\Sigma}$ depends 
on the choice of gauge).
The spatial average and the hypersurface integral of 
$q (x)$, are related by
\be\label{hdx}
\  \langle q \rangle = \langle \langle q
\ ({g^{(3)}/g^B})^{\fr{1}{2}} \rangle \rangle,
\ee
where we used the definition (\ref{hypint}).
The integrand at the right-hand side of equation (\ref{hdx}) is gauge dependent,
and may be expanded in powers of $\lambda$ about
$\lambda =0$, i.e.,
\be\label{esv}
\ q  ({g^{(3)}/g^B})^{\fr{1}{2}} (\lambda,\bar{p}) =
\ \sum_{k=0}^{k=\infty} \fr{\lambda^k}{k!} {\cal L}^k_{\xi}
\ q  ({g^{(3)}/g^B})^{\fr{1}{2}} ,
\ee
where ${\cal L}^k_{\xi}$ denotes the $k$-th order Lie derivative
with respect to $\xi$, evaluated in $\bar{p}$.
By substituting the expansion (\ref{esv}) in the integrand
at the right-hand side of equation (\ref{hdx}), we obtain
\be\label{vcn}
\  \langle q \rangle (\lambda) - \langle q \rangle (0)  =
\  \sum_{k=1}^{k=\infty}  \fr{\lambda^k}{k!}
\ \langle \langle{\cal L}^k_{\xi}  q ({g^{(3)}/g^B})^{\fr{1}{2}} \rangle \rangle.
\ee
For $k=1$, the contribution to the right-hand side of equation (\ref{vcn})
is evaluated using
\be\label{nd}
\ {\cal L}_{\xi} q = \xi^i q_{;i}
\ee
and
\be\label{ytr}
\ {\cal L}_{\xi} ({g^{(3)}/g^B})^{\fr{1}{2}} = \xi^i_{;i} ({g^{(3)}/g^B})^{\fr{1}
{2}},
\ee
where $;$ denotes covariant differentiation with respect to $g_{ij}$.
Combining equations (\ref{nd}) and (\ref{ytr}) yields,
\be\label{ncd}
\  {\cal L}_{\xi}  q ({g^{(3)}/g^B})^{\fr{1}{2}}  =
\  (q \xi^i)_{;i} ({g^{(3)}/g^B})^{\fr{1}{2}} ,
\ee
and
\be
\ \langle \langle (q \xi^i)_{;i} ({g^{(3)}/g^B})^{\fr{1}{2}} \rangle \rangle = 0,
\ee
due to Gauss's theorem.
The $k=2$ contribution to the right-hand side of equation
(\ref{vcn}) is obtained by making the substitution
$q \rightarrow (q \xi^i)_{;i}$ in expression (\ref{ncd}),
and for arbitrary $k \in {\bf Z}^+$ the same result
follows by induction.
Since the terms at the right-hand side of equation (\ref{vcn}) vanish
for all $k$, we established that the spatial average of a scalar
function $q$ is gauge invariant to arbitrary order in the
expansion parameter $\lambda$.
Applying the same analysis as above to the hypersurface integral
of a scalar field $q (x)$, we find,
\be\label{vcnn}
\  \langle \langle q \rangle \rangle (\lambda)
\ - \langle \langle q \rangle \rangle  (0)  =
\  \sum_{k=1}^{k=\infty}  \fr{\lambda^k}{k!}
\ \langle \langle {\cal L}^k_{\xi}  q  \rangle \rangle,
\ee
which depends on $\lambda$, due to equation (\ref{nd}),
unless $q$ is a constant on $\Sigma$.
A similar derivation, where we reverse the roles of the  
spacetimes $\Sigma$ and $\bar{\Sigma}$, shows that  
the hypersurface integral of a scalar field $\bar{q}(\bar{x})$ 
which lives in $\bar{\Sigma}$ is gauge invariant,
while the spatial average of $\bar{q}(\bar{x})$ is
gauge invariant iff $\bar{q}(\bar{x})$ is
constant in $\bar{\Sigma}$.
It follows from these observations that the spatial
average of a perturbation $\delta q := q (x) - \bar{q}(\bar{x})$
is gauge invariant iff $\bar{q}(\bar{x})$ is constant on 
$\bar{\Sigma}$.

Note that Futamase in  \cite{futamase}  uses the
hypersurface integral as a spatial averaging operation in
the calculation of second order effects, while he does not
consistently fix a gauge in these papers (namely, he assumes
a comoving synchronous gauge, {\em and} constant
expansion on the hypersurfaces of constant time coordinate).

\section{Appendix B}
 
In this appendix we derive the decomposition of the
generic linearized averaging operation
\be\label{cc2}
\ \hat{A}^{\infty} := \lim_{n \rightarrow \infty}
\ \hat{A}^{(1) n} \delta g_{\mu \nu} ,
\ee
in terms of the spatial average $\langle \delta g_{\mu \nu} \rangle (t)$,
which is uniquely defined.
Note that the existence of the limit (\ref{cc2}) implies that
\be\label{ccs}
\ \hat{A}^{(1)} \delta g^*_{\mu\nu}  = \delta g^*_{\mu\nu}
\ee
for arbitrary spatially homogeneous and isotropic
perturbations $\delta g^*_{\mu\nu}$
(up to the freedom of diffeomorphisms acting
at either side of equation (\ref{ccs})).
 
Without loss of generality, a spatially
homogeneous and isotropic perturbation
$\delta g^*_{\mu\nu}$ about $\bar{\bf S}$
can be written in the form
\be\label{g*}
\ \delta g^*_{\mu\nu}
\ = \phi_1 (\bar{t}) \bar{n}_{\mu} \bar{n}_{\nu} + \phi_2 (\bar{t}) \bar{h}_{\mu\nu},
\ee
where $\bar{h}_{\mu\nu} := {g}^B_{\mu\nu} +  \bar{n}_{\mu} \bar{n}_{\nu}$,
and $\bar{n}_{\mu}$ denotes the timelike
future directed vector in $\bar{\bf S}$ which is orthogonal
to $\bar{\Sigma}$, and which is normalized with respect to the
background metric $ {g}^B_{\mu\nu}$, and $\phi_1 (\bar{t})$
and $\phi_2 (\bar{t})$ are arbitrary functions of $\bar{t}$.
 
When we substitute expression (\ref{g*}) for $\delta g^*_{\mu\nu}$
and expression (\ref{spav}) for $\hat{A}^{(1)}$
into condition (\ref{ccs}), we obtain,
\[
\ \int {\it d} t^{\pr} {\it d}^3 x^{\pr} \left(
\ \phi_1 (t^{\pr})  \bar{n}_{\rho} \bar{n}_{\sigma}
\ + \phi_2 (t^{\pr})  \bar{h}_{\rho\sigma}\right)
\ f^{\rho\sigma}_{\mu\nu} (x,x^{\pr})
\]
\be\label{ind}
\ = \phi_1 (t) \bar{n}_{\mu} (x) \bar{n}_{\nu} (x)   + \phi_2 (t) \bar{h}_{\mu\nu} (x),
\ee
for arbitrary functions $\phi_1 (t)$ and $\phi_2 (t)$.
Equation (\ref{ind}) holds for arbitrary $\phi_1 (t)$
and $\phi_2 (t)$ iff
\be\label{fdi}
\ \int_{\bar{\Sigma}} {\it d}^3 x^{\pr}
\ \bar{n}_{\rho} (x^{\pr}) \bar{n}_{\sigma} (x^{\pr})
\ f^{\rho\sigma}_{\mu\nu} (x,x^{\pr})
\ = \delta (t^{\pr} - t) \bar{n}_{\mu}(x) \bar{n}_{\nu}(x),
\ee
and
\be\label{fdy}
\ \int_{\bar{\Sigma}} {\it d}^3 x^{\pr}
\ \bar{h}_{\rho\sigma} (x^{\pr})
\ f^{\rho\sigma}_{\mu\nu} (x,x^{\pr})
\ = \delta (t^{\pr} - t) \bar{h}_{\mu\nu} (x).
\ee
Expression (\ref{fdi}) shows that
$f^{\rho\sigma}_{\mu\nu} (x,x^{\pr})$ is proportional
to a delta distribution $\delta (t - t^{\pr})$. It follows from
this observation that $\hat{A}^{(1)}$ can be naturally
defined in terms of a linearized
{\em spatial} averaging operation
$\hat{A}^{(1)}_s$, i.e.,
\be\label{kkkt}
\ \hat{A}^{(1)} \delta g_{\mu\nu}  
\ = \hat{A}^{(1)}_s (t) \delta g_{\mu\nu},
\ee
where $\hat{A}^{(1)}_s $ is defined by,
\be\label{spatiav}
\ \hat{A}^{(1)}_s  \delta g_{\mu\nu}  =
\    \int_{\bar{\Sigma} (t)} {\it d}^3 x^{\pr}
\ f^{\rho\sigma}_{\mu\nu} (t,x^i,x^{i \pr})
\ \delta g_{\rho\sigma} (x^{i \pr}) ,
\ee
and
\be
\ f^{\rho\sigma}_{\mu\nu} (t,x^i,x^{i \pr}) :=
\ \int_{\Delta t^{\pr}} {\it d} t^{\pr}
\ f^{\rho\sigma}_{\mu\nu} (t,t^{\pr},x^i,x^{i \pr}),
\ee
and ${\Delta t^{\pr}}$ is chosen such that $t \in {\Delta t^{\pr}}$.
At first sight, the decomposition of the linear
averaging operation $\hat{A}^{(1)}$ in terms of a spatial
averaging operation which is defined on a collection of
spatial hypersurfaces might be surprising, since the
choice of a collection of spatial hypersurfaces
$\bar{\Sigma} (t)$ in ${\bf S}$ is gauge dependent.
It will be shown in section (\ref{gp}) that although the
choice of $\bar{\Sigma} (t)$ in ${\bf S}$ is gauge dependent, the
linearized spatial averaging operation (\ref{kkkt}) is
to first order gauge independent.
 
Assuming that the limit (\ref{c2}) exists, then
by substituting expression (\ref{kkkt}) into expression (\ref{cc2})
one finds that the limit
\be\label{hdd}
\ \langle \delta g_{\mu\nu} \rangle := \lim_{n \rightarrow \infty}
\ \hat{A}^{(1) n}_s \delta g_{\mu\nu},
\ee
exists.
We will show that the limiting spatial averaging
operation which is defined by equation (\ref{hdd}) is universal.
 
Expression (\ref{spatiav}) and the definition
(\ref{hdd})  imply that,
\be\label{bbs}
\ \langle \delta g_{\mu\nu} \rangle := \lim_{n \rightarrow \infty}
\ \int_{\bar{\Sigma} (t)} {\it d}^3 x^{\pr}
\  f^{ n \rho\sigma}_{\mu\nu} (t,x^i,x^{\pr i})
\ \delta g_{\rho\sigma} (x^{\pr i}) ,
\ee
where $f^{n \rho\sigma}_{\mu\nu}$ is defined in terms of
$f^{\rho\sigma}_{\mu\nu}$ by
induction over $n$,
\be\label{klj}
\ f^{ n \rho\sigma}_{\mu\nu} (x^i,x^{\pr i}) = \int_{\bar{\Sigma} (t)}
\ {\it d}^3 q  f^{\rho\sigma}_{\alpha\beta} (t,x^{\pr i}, q^i)
\ f^{ n-1 \alpha\beta}_{\mu\nu} (t,x^i, q^i),
\ee
and $f^{ 1 \rho\sigma}_{\mu\nu} := f^{ \rho\sigma}_{\mu\nu}$.
Let us now try to determine the limit
\be\label{ddf}
\ f^{\infty \rho\sigma}_{\mu\nu}
\ := \lim_{n \rar \infty} f^{n \rho\sigma}_{\mu\nu}.
\ee
An explicit calculation of $f^{\infty \rho\sigma}_{\mu\nu}$,
using the definition definition (\ref{klj})
for $f^{n \rho\sigma}_{\mu\nu}$ and starting with arbitrary
realizations for $f^{\rho\sigma}_{\mu\nu}$, would be quite
cumbersome, but fortunately it appears that the symmetries
of the background spacetime $\bar{S}$, and the stability
condition (\ref{c2}) determine
$f^{\infty \rho\sigma}_{\alpha\beta}$
completely.
 
Recall that we required that the limit (\ref{c2}) converges
to a spatially homogeneous and isotropic metric perturbation
for arbitrary perturbations
$\delta g_{\mu\nu}$, which implies that
\be\label{qqs}
\ \langle \delta g_{\mu\nu} \rangle (x^i) =
\ \int_{\bar{\Sigma} (t)} {\it d}^3 x^{\pr}
\  f^{ \infty \rho\sigma}_{\mu\nu} (t,x^i,x^{\pr i})
\ \delta g_{\rho\sigma} (x^{\pr i})  = \delta g^*_{\mu\nu},
\ee
for all $x$, where we used expression (\ref{kkkt}) and
$\delta g^*_{\mu\nu}$ has the form (\ref{g*}).
If expression (\ref{qqs}) holds for arbitrary perturbations
$\delta g_{\rho\sigma} (x^{\pr i})$, it also holds for
arbitrary perturbations
$\delta g_{\rho\sigma} (x^{\pr i} + c^i)$,
where $c^i \in {\bf R}$.
By absorbing the constants $c^i$ into the coordinates $x^i$, one
finds that expression (\ref{qqs}) remains unchanged under
the substitution
\be\label{trinv1}
\ f^{ \infty \rho\sigma}_{\mu\nu} (x^i,x^{\pr i}) \rar
\ f^{ \infty \rho\sigma}_{\mu\nu} (x^i,x^{\pr i} - c^i ).
\ee
Furthermore, since the right-hand side of equation (\ref{qqs})
is spatially homogeneous by requirement, we find that the
left-hand side of equation (\ref{qqs}) must be also
invariant under the
substitution
\be\label{trinv2}
\ f^{ \infty \rho\sigma}_{\mu\nu} (x^i,x^{\pr i}) \rar
\ f^{ \infty \rho\sigma}_{\mu\nu} (x^i + d^i,x^{\pr i} ),
\ee
where $d^i \in {\bf R}$ is arbitrary.
Since equation (\ref{qqs}) is invariant under
(\ref{trinv1}) and (\ref{trinv2}) for arbitrary
perturbations
$\delta g_{\mu\nu}$, we conclude that
$f^{ \infty \rho\sigma}_{\mu\nu}$
is (up to the freedom of performing diffeomorphisms)
constant on $\Sigma$ when regarded as a
distribution (i.e., neglecting sets
of Lebesque measure zero).
Furthermore, since equation (\ref{qqs}) holds for arbitrary
$\delta g_{\mu\nu}$, the distribution
$f^{ \infty \rho\sigma}_{\mu\nu} (x,x^{\pr})$ must
be proportional to a tensor of the form (\ref{g*})
in the point $x$, thereby fixing the $\mu\nu$ dependent
part of $f^{ \infty \rho\sigma}_{\mu\nu}$.
We may therefore write
\be\label{dsa}
\ f^{ \infty \rho\sigma}_{\mu\nu} (x,x^{\pr})
\ =  g_1^{ \rho\sigma} (x^{\pr}) \bar{n}_{\mu} (x) \bar{n}_{\nu} (x)
\ +  g_2^{ \rho\sigma} (x^{\pr}) \bar{h}_{\mu\nu} (x) ,
\ee
where $g_1^{ \rho\sigma} (x^{\pr})$ and
$g_2^{ \rho\sigma} (x^{\pr})$ are spatially homogeneous tensor
densities in $x^{\pr}$, and we used expression
(\ref{g*}) for $g^*_{\mu\nu}$.
 
A similar argument, using the invariance of equation (\ref{qqs})
under the group of spatial rotations (using that ${\cal F}$ does
not explicitly
depend on $x$), shows that the bi-tensor density
$f^{ \infty \rho\sigma}_{\mu\nu} (x,x^{\pr})$ is isotropic
with respect to the indices $\sigma$ and $\rho$, which
implies that the tensor densities $g_1^{ \rho\sigma} (x^{\pr})$
and $g_2^{ \rho\sigma} (x^{\pr})$ in expression (\ref{dsa})
are of the form,
\be\label{alpha}
\ g_1^{ \rho\sigma} (x^{\pr}) = \alpha_1 \sqrt{{g}^{(3)}}
\ \bar{n}^{\rho} (x^{\pr}) \bar{n}^{\sigma} (x^{\pr})
\ + \alpha_2  \sqrt{{g}^{(3)}} \bar{h}^{\rho\sigma} (x^{\pr})
\ee
and
\be\label{beta}
\ g_2^{ \rho\sigma} (x^{\pr}) = \alpha_3 \sqrt{{g}^{(3)}}
\  \bar{n}^{\rho} (x^{\pr}) \bar{n}^{\sigma} (x^{\pr})
\ + \alpha_4 \sqrt{{g}^{(3)}}  \bar{h}^{\rho\sigma} (x^{\pr})
\ee
where ${g}^{(3)}$ denotes the real space volume element, which
follows from requiring spatial gauge invariance
at higher orders (see appendix A), and the factors $\alpha_n$
($n \in \{1,2,3,4 \}$) are constant on $\Sigma$.
Substituting expressions (\ref{alpha}) and (\ref{beta}) in expression
(\ref{dsa}) yields
\be\label{dsaX}
\ f^{ \infty \rho\sigma}_{\mu\nu} (x,x^{\pr})
\ = \sqrt{{g}^{(3)}} ( \alpha_1  \bar{n}^{\rho} (x^{\pr}) \bar{n}^{\sigma} (x^{\pr})
\ \bar{n}_{\mu} (x) \bar{n}_{\nu} (x)
\   +  \alpha_4  \bar{h}^{\rho\sigma} (x^{\pr}) \bar{h}_{\mu\nu} (x) ),
\ee
where we used expressions (\ref{fdi}) and (\ref{fdy}) to show
that the terms proportional to $\alpha_2$ and $\alpha_3$
vanish.
 
By substituting expression (\ref{dsaX}) for $f^{ \infty \rho\sigma}_{\mu\nu}$
into condition (\ref{qqs}), where we set
$\delta g_{\rho\sigma}$ equal to $\delta g^*_{\mu\nu}$ defined
by expression (\ref{g*}), we find that the constants $\alpha_1$ and $\alpha_4$
in expression (\ref{dsaX}) must satisfy the condition
\be\label{hs}
\ \int_{\bar{\Sigma} (t)} {\it d}^3 x^{\pr} \sqrt{{g}^{(3)}} \alpha_1 =
\ 3 \int_{\bar{\Sigma} (t)} {\it d}^3 x^{\pr} \sqrt{{g}^{(3)}} {\alpha_4} =1.
\ee
Expression (\ref{hs}) shows that the constants $\alpha_1$ and
$3 \alpha_4$ are equal to (volume$({\Sigma}))^{-1}$
when ${\Sigma}$ is closed, while in the case when ${\Sigma}$
is open, $\alpha_1$ and $\alpha_4$ are defined in a distributional
sense by condition (\ref{hs}), and by the condition that
$\alpha_1$ and $\alpha_4$
are constant on ${\Sigma} (t)$.
 
By substituting expression (\ref{dsaX}) into expression (\ref{qqs})
we obtain the explicit expression for the spatial average,
\be\label{bcs}
\ \langle \delta g_{\mu\nu} \rangle =
\ \int_{\bar{\Sigma} (t)} {\it d}^3 x^{\pr} \sqrt{{g}^{(3)}} \alpha_1
\ee
\be
\ [ \bar{n}^{\rho} (x^{\pr}) \bar{n}^{\sigma} (x^{\pr})
\   \bar{n}_{\mu} (x) \bar{n}_{\nu} (x)
\ + \fr{1}{3} \bar{h}^{\rho\sigma} (x^{\pr}) \bar{h}_{\mu\nu} (x) ]
\ \delta g_{\rho\sigma}.
\ee
Note that $ \bar{n}^{\rho} \bar{n}^{\sigma}  \delta g_{\rho\sigma}$
equals the perturbation of $g_{00}$ in coordinates which are synchronous
in the background (i.e., coordinates for which
$ {g}^B_{\mu 0} = - \delta_{\mu}^0$),
while $\bar{h}^{\rho\sigma}\delta g_{\rho\sigma}$ equals the perturbation
of the spatial volume element on $\Sigma$, to first
order.
 
Summarizing the derivation in this appendix, we showed
that the general linearized averaging operation which is a
functional of metric perturbations about FLRW, and for
which unperturbed FLRW is a stable fixed point, has a
unique limit when applied iteratively to perturbed FLRW.
Furthermore, we showed that this linearized averaged operation
is naturally defined in terms of a spatial averaging operation
which works on $g_{00}$ and the spatial volume perturbation
in coordinates which are synchronous in the background.

\end{document}